\documentclass[final]{article}
\usepackage{graphicx}
\usepackage{rotating}
\usepackage{amssymb}
\usepackage{mathptmx}
\usepackage[numbers]{natbib}
\usepackage{subscript}
\usepackage{gensymb} 
\usepackage{fancyhdr}
\usepackage{lastpage}
\usepackage{abstract}
\usepackage{hyphenat}
\usepackage{marvosym}
\usepackage{ragged2e}
\usepackage{subscript}
\usepackage{booktabs}
\usepackage{tabularx}
\usepackage{xcolor}

\usepackage[vcentering,dvips,left=3.2cm,right=3.2cm,top=4cm,bottom=4cm]{geometry}

\usepackage{authblk}

\usepackage{fancyhdr}
\renewenvironment{abstract}{%
  \hfill\par
    {\noindent\bfseries Abstract\par}
    \noindent\rule{\textwidth}{1pt}}
    {\par\noindent\rule{\textwidth}{1pt}
}

\makeatletter

\renewcommand\@maketitle{%
  \hfill
  \vskip 2em
  \let\footnote\thanks 
  {\noindent\LARGE \bfseries\@title \par }
  \vskip 1.5em
  {\noindent\large\@author\par}
  \vskip 1em \par
}
\makeatother

\newenvironment{acknowledgements}
    {\par\addvspace{17pt}\small\rmfamily\noindent {\bfseries Acknowledgements}%
    \small\rmfamily\noindent}%
    {}%

\makeatletter

\bibpunct{}{}{,}{s}{}{,}

\begin{document}

\newcommand{\hdblarrow}{H\makebox[0.9ex][l]{$\downdownarrows$}-}
\title{\nohyphens{Development of Microwave Superconducting Microresonators for Neutrino Mass Measurement in the HOLMES Framework}}

\author[1]{A.Giachero\footnote{\Telefon~+39 02-6448-2456, \Letter~Andrea.Giachero@mib.infn.it}}
\author[3]{P.~Day}
\author[4]{P.~Falferi}
\author[2,1]{M.~Faverzani}
\author[2,1]{E.~Ferri}
\author[2]{C.~Giordano}
\author[5]{B.~Margesin}
\author[6]{R.~Mezzena}
\author[2]{R.~Nizzolo}
\author[2,1]{A.~Nucciotti}
\author[2,1]{A.~Puiu}
\author[2]{L.~Zanetti}
\affil[1]{INFN Milano-Bicocca, Milano, Italy}
\affil[2]{Universit\`a di Milano-Bicocca, Milano, Italy}
\affil[3]{Jet Propulsion Laboratory, Pasadena, CA, U.S.A.}
\affil[4]{Istituto di Fotonica e Nanotecnologie, CNR-Fondazione Bruno Kessler, Trento, Italy}
\affil[5]{Fondazione Bruno Kessler, Trento, Italy}
\affil[6]{Dipartimento di Fisica, Universit\`{a}  di Trento, Trento, Italy}


\maketitle

\justify\begin{abstract}
The European Research Council has recently funded HOLMES, a project with the aim of performing a calorimetric measurement of the electron neutrino mass measuring the energy released in the electron capture decay of \textsuperscript{163}Ho. The baseline for HOLMES are microcalorimeters coupled to Transition Edge Sensors (TESs) read-out with rf-SQUIDs, for microwave multiplexing purposes. A promising alternative solution is based on superconducting microwave resonators, that have undergone rapid development in the last decade. These detectors, called Microwave Kinetic Inductance Detectors (MKIDs), are inherently multiplexed in the frequency domain and suitable for even larger-scale pixel arrays, with theoretical high energy resolution and fast response. The aim of our activity is to develop arrays of microresonator detectors for X-ray spectroscopy and suitable for the calorimetric measurement of the energy spectra of \textsuperscript{163}Ho. Superconductive multilayer films composed by a sequence of pure Titanium and stoichiometric TiN layers show many ideal properties for MKIDs, such as low loss, large sheet resistance, large kinetic inductance, and tunable critical temperature $T_c$. We developed Ti/TiN multilayer microresonators with $T_c$ within the range from 70\,mK to 4.5\,K and with good uniformity. In this contribution, we present the design solutions adopted, the fabrication processes, and the characterization results.
\end{abstract}

\section{Introduction}
The HOLMES~\cite{HOLMES} experiment is aimed at directly measuring the electron neutrino mass using the Electron Capture (EC) decay of \textsuperscript{163}Ho, as proposed in 1982 by De Rujula and Lusignoli~\cite{DeRujula}. In order to push down the current sensitivity on the neutrino mass, HOLMES will deploy large detector arrays of low-temperature microcalorimeters implanted with \textsuperscript{163}Ho nuclei. The baseline for HOLMES are Mo:Cu TESs (Transition Edge Sensors) on SiN\textsubscript{x} membrane with Gold absorbers, read-out with a rf-SQUID, for microwave multiplexing purposes. 

In addition to the HOLMES experiment, there are currently two other experimental activities aiming at measuring the neutrino mass using the EC decay of \textsuperscript{163}Ho: NuMECS~\cite{NuMECS} and ECHo~\cite{ECHo}. The former uses arrays of TESs, while the latter uses array of Metallic Magnetic Calorimeters (MMCs). MMC represents an alternative technology to TES with good energy resolution and very fast signal rise time~\cite{Enss}.

\begin{figure}[!t]
\begin{minipage}[r]{.50\textwidth}
  \centering
  \includegraphics[width=\textwidth]{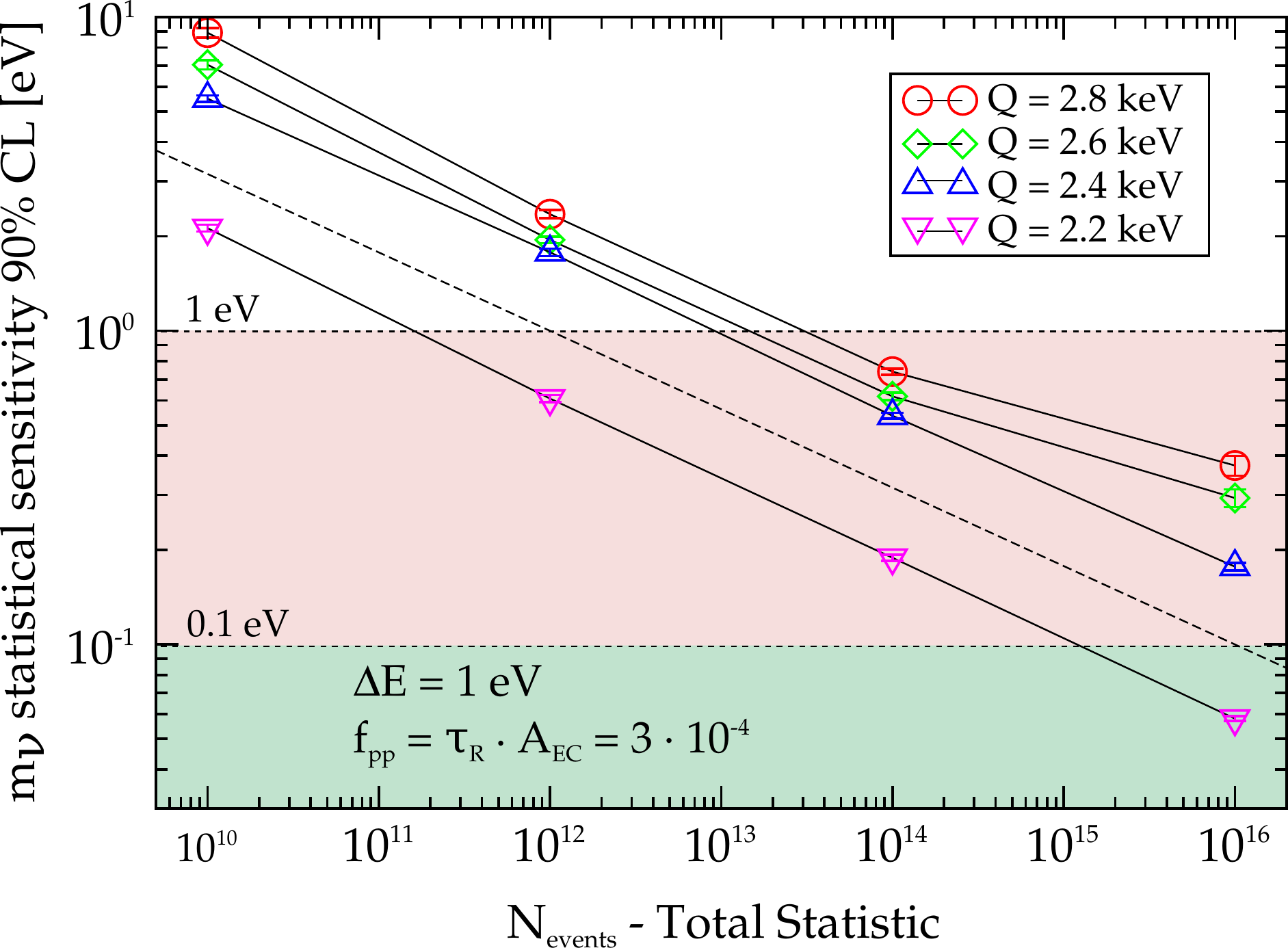}
\end{minipage}
\begin{minipage}[r]{.50\textwidth}
  \centering
  \scalebox{0.85}{
  \begin{tabular}{ccccc}
  \toprule
  $A_{EC}$ & $\tau_{rise}$ & $\Delta E$ & $N_{ev}$ & Exposure \\
  {[Hz]} & [$\mu$s]    & [eV]       & [counts] & [det$\cdot$year]\\
  \midrule
  1   & 0.1 & 0.3 & $1.2\cdot 10^{14}$ & $3.9\cdot 10^{6}$\\
  \textcolor{red}{100} & \textcolor{red}{0.1} & \textcolor{red}{0.3} & \textcolor{red}{$6.4\cdot 10^{14}$} & \textcolor{red}{$2.0\cdot 10^{5}$}\\
  100 & 0.1 & 1   & $7.4\cdot 10^{14}$ & $2.4\cdot 10^{5}$\\
  10  & 0.1 & 1   & $4.5\cdot 10^{14}$ & $1.5\cdot 10^{6}$\\
  10  & 1   & 1   & $7.4\cdot 10^{14}$ & $2.4\cdot 10^{6}$\\
  \bottomrule
  \end{tabular}}
\end{minipage}
\caption{(Left) \textsuperscript{163}Ho decay experiments statistical sensitivity dependence on the total statistics $N_{events}$. (Right) Exposure required for $m_\nu=0.1$\,eV sensitivity and  $Q_{EC}=2800$\,eV. Figure and table reprinted from A. Nucciotti~\cite{NUCCIOTTI}.} 
\label{fig:sensitivity}
\end{figure}
Recent high-precision Penning-trap measurements~\cite{PENNING} set the \textsuperscript{163}Ho EC Q-value at 2833\,eV. This means for HOLMES a 90\% confidence level statistical neutrino mass sensitivity greater than 1\,eV~\cite{NUCCIOTTI}. Considering the simulation results reported in figure~\ref{fig:sensitivity}, a possible experimental approach, beyond HOLMES, with the aim to reach the sub-eV sensitivity ($m_\nu<0.1$\,eV), requires an experiment composed by 6 microcalorimeter arrays, each one with $10^6$\,pixels, with a very fast detector response ($\tau <1\,\mu$s), running for 10\,years, for a total number of \textsuperscript{163}Ho nuclei around $8\cdot 10^{19}$\,nuclei. To achieve these requirements, the multiplexing factor and the read-out bandwidth play a crucial role for future developments. 

TESs with SQUID read-out demonstrate their scalability to large arrays in several applications by using different multiplexing techniques: Time Division (TDM)~\cite{TDM}, Code Division (CDM)~\cite{CDM}, and Frequency Division (FDM)~\cite{FDM}. However, the scalability of these readout approaches is limited by the finite measurement bandwidth ($\simeq$10\,MHz) achievable in a flux-locked loop. Considering an event rate of 300\,Hz expected for HOLMES, a larger bandwidth is needed in order to resolve possible pile-up. To fulfill this requirement, HOLMES will exploit recent advancements on microwave multiplexing ($\mu$MUX). This techniques is based on the use of rf-SQUIDs as input devices with flux ramp modulation\cite{RFSQUID}. Despite this novel approach is very promising and will guarantee the multiplexing factor needed for HOLMES (1000 pixels), todate there is no practical demonstration of the scalability to Mega-pixel arrays needed for a sub-eV mass sensitivity measurement.  
   
An alternative solution to realize large-pixel arrays with a large-multiplexing factor is based on MKIDs detectors. The natural multiplexing capability of a MKID in the frequency domain allows to read-out up to thousand detectors using a single pair of coaxial cables and a single HEMT amplifier. Comparing the same number of pixels of rf-SQUID coupled TESs, MKIDs use simpler cryogenics and read-out schemes since they do not need the bias and the ramp modulation lines. Besides, MKIDs do not require the use of SQUIDs, and are built as single or few layers of superconducting thin film, simplifying the fabrication process.  MKIDs have demonstrated their scalability in several experiments. The running experiment ARCONS employs 2024 MKID pixels~\cite{ARCONS} and future experiments, as GIGA-Z\cite{GIGAZ} and KRAKENS~\cite{KRAKENS}, will employ up to 100\,kpixel arrays. 


\section{MKIDs for X-ray Spectroscopy}
MKIDs are a cryogenic detector technology proposed for the first time in 2003~\cite{Day}. Traditionally, MKIDs have been thought of as non-equilibrium detectors, which detect the excess of quasiparticles from the absorbed photon (\textit{athermal mode}). They have demonstrated their scalability operating in this way in several instruments for sub-mm radiometry. MKIDs in athermal mode can also be used for X-ray spectroscopy by electrically coupling a superconducting absorber to the microresonator. The absorber must have a high stopping power (high Z) and an energy gap greater than the gap of the material constituting the resonator, in order to exploit the quasiparticle trapping effect~\cite{Tantalum} and avoid significant energy loss. Tantalum represents a good choice. The energy resolution is set by the statistical fluctuations in the initial number of quasiparticle and, considering a Tantalum absorber $200\times 35\,\mu$m$^2$ wide and 600\,nm thick, the theoretical value is around 3\,eV\, at 6\,keV.

The \textit{thermal mode} is a variation on the MKID classical way of operation that has generated interest in recent years. As demonstrated in Gao \textit{et al.}~\cite{ThermalEqui}, a temperature change can produce an identical increase of quasiparticle population of an external pair-breaking source. Exploiting this behavior, the so-called TKIDs (Thermal Kinetic Inductance Detectors) can be used as thermometers to detect the temperature rise due to a X-ray by exploiting an absorber thermally coupled with the microresonator (pure calorimeters). As for TESs, the energy resolution is theoretically limited only by thermodynamic fluctuations across the thermal weak links: $\Delta E_{FWHM} \simeq 2.355\,\sqrt{kT^2 C}$, where $k$ is the Boltzmann constant, $C$ the heat capacity of the absorber, and $T$ the bath temperature. Considering the intrinsic properties of a metal absorber (i.e. Gold) $200\times 200\,\mu$m$^2$ wide and 2\,$\mu$m thick working at $T = 50$\,mK, it is possible to have resolution around 1\,eV at 6\,keV. A recent work demonstrated an encouraging preliminary energy resolution of 75\,eV at 6\,keV~\cite{TKID}.  

For both the operating modes, the time resolution depends on the resonance frequency ($f_r$) and quality factor ($Q$) of the resonators. Designing a quality factor around $Q\simeq 10^4$ corresponds to a resonator response time of $\tau_r = Q/\pi f_0$ which sets the rise time of the detector. Considering individual microresonator frequencies distributed in the $(1\div 6)$~GHz range, rise time around $1\mu $s or less is achievable. 
  
The goal of our R\&D activity is to realize superconducting microresonator arrays with energy and time resolution around 1\,eV and 1\,$\mu$s respectively, suitable for next generation neutrino mass measurement experiments. To achieve this goal, we are studying and testing different design geometries, and we are optimizing the sensor fabrication. To date we have produced two different families of chips with two different critical temperatures: films with \textit{high} $T_c$ to study the athermal mode and films with \textit{low} $T_c$ to study the thermal mode. As high $T_c$ we targeted a value around 1.5\,K in order to have slower recombination time, with respect stoichiometric TiN film ($T_c=4.5$\,K), and a low energy gap, more than 3 times less than the Tantalum one ($\Delta =0.67$\,meV). As low $T_c$ we targeted a value around 500\,mK in order to maximize the variation of the resonance frequency shift $\Delta f_r$ as a function of the temperature.

In this work, we present a preliminary production of Ti/TiN multilayer microresonator $2\times 8$ pixel arrays with not implemented absorbers. Since it was the first time we tried to realize multilayers with very low critical temperatures, the aim of this fabrication run was to validate the production processes and the correct behavior as resonators. Superconducting microresonators without absorber are not able to resolve a monochromatic energy, as also reported in our previous work~\cite{Faverzani_LTD15}. For this reason, we present here the characterization of all the relevant parameters but the energy resolution.

\section{Materials and Fabrication}
\begin{figure}[!t] 
 \begin{center}
    \includegraphics[clip=true,width=0.8\textwidth]{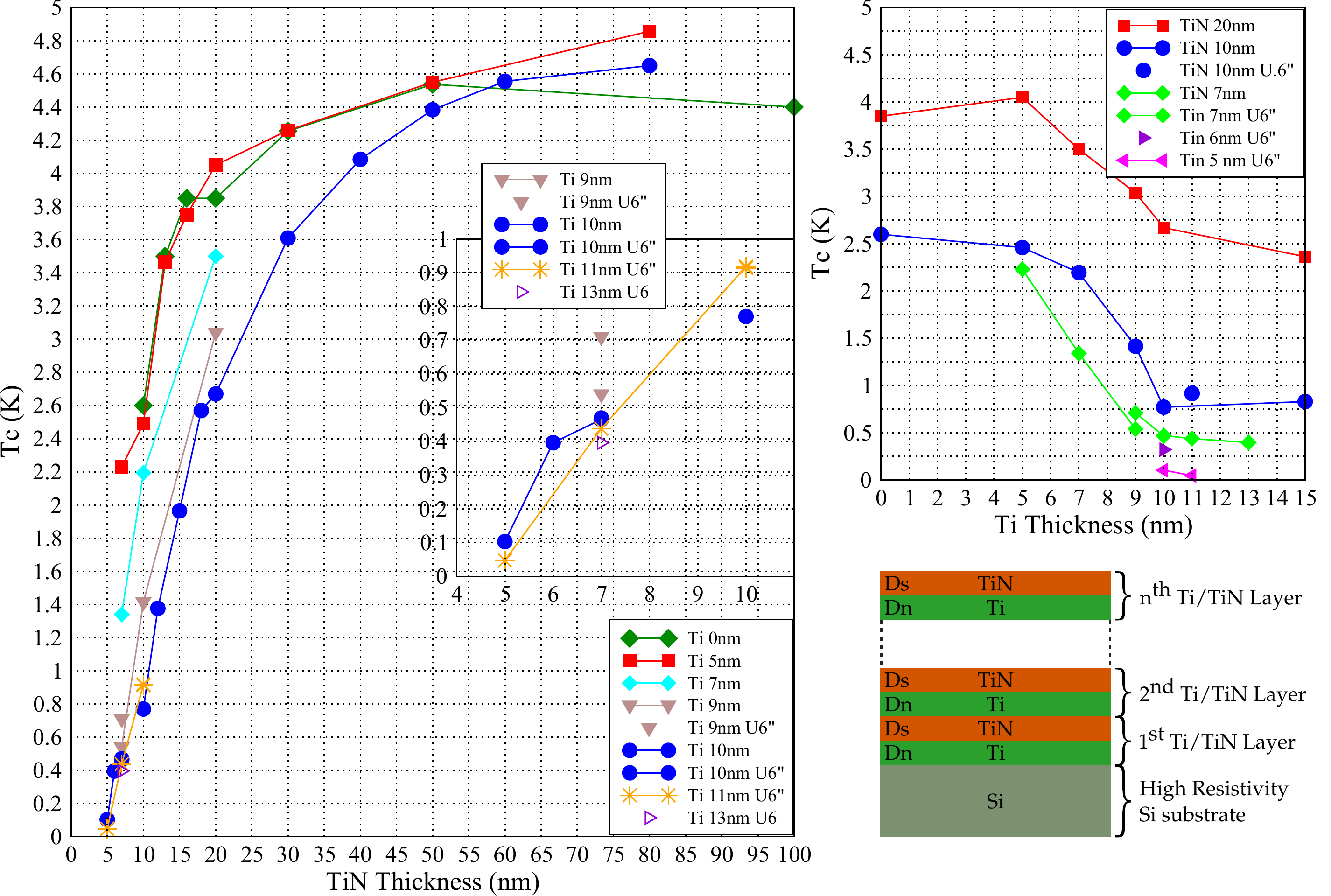}
  \end{center}
  \caption{\label{fig:TiN} Plots show the measured $T_c$ of the Ti/TiN multilayers as function of TiN (left) and Ti (right) single-layers thicknesses. In the bottom-right, the schematic of multilayers is also reported. A good reproducibility and uniformity in the films was obtained, from the edge and from the center the critical temperature varied by no more than 1\%.}
\end{figure}

The main important requirements for our superconducting films are a very low transition temperature ($T_c\simeq 500$~mK) for the \textit{thermal mode} and an energy gap 3 - 4 times lower than the Tantalum energy gap for the \textit{athermal mode}. These requirements can be satisfied by employing microresonators built in Ti/TiN multilayer technology~\cite{Multilayer}. In fact, exploiting the proximity effect~\cite{PROXIMITY} the $T_c$ of a superconducting material can be reduced by the superposition of a normal metal or a metal with a superconductive transition at lower temperature. In this configuration, the Cooper pairs leak into the normal metal; consequentially the pair density in the superconducting metal is reduced as well as the transition temperature. The magnitude of this effect depends on the thickness of the two layers. In the case of Ti/TiN multilayers, by adjusting these thicknesses, it is possible to tune the $T_c$ between the transition temperature of Ti (0.4~K) and TiN (4.5~K). Results shown in figure \ref{fig:TiN} prove that the critical temperature can actually be tuned within the range from 70\,mK to 4.5\,K by properly choosing the Ti thickness in the ($0-15$)~nm range and the TiN thickness in the ($5-100$)~nm range with a good reproducibility and uniformity.

\section{Characterization Measurements}
\begin{figure}[!t] 
 \begin{center}
    \includegraphics[clip=true,width=\textwidth]{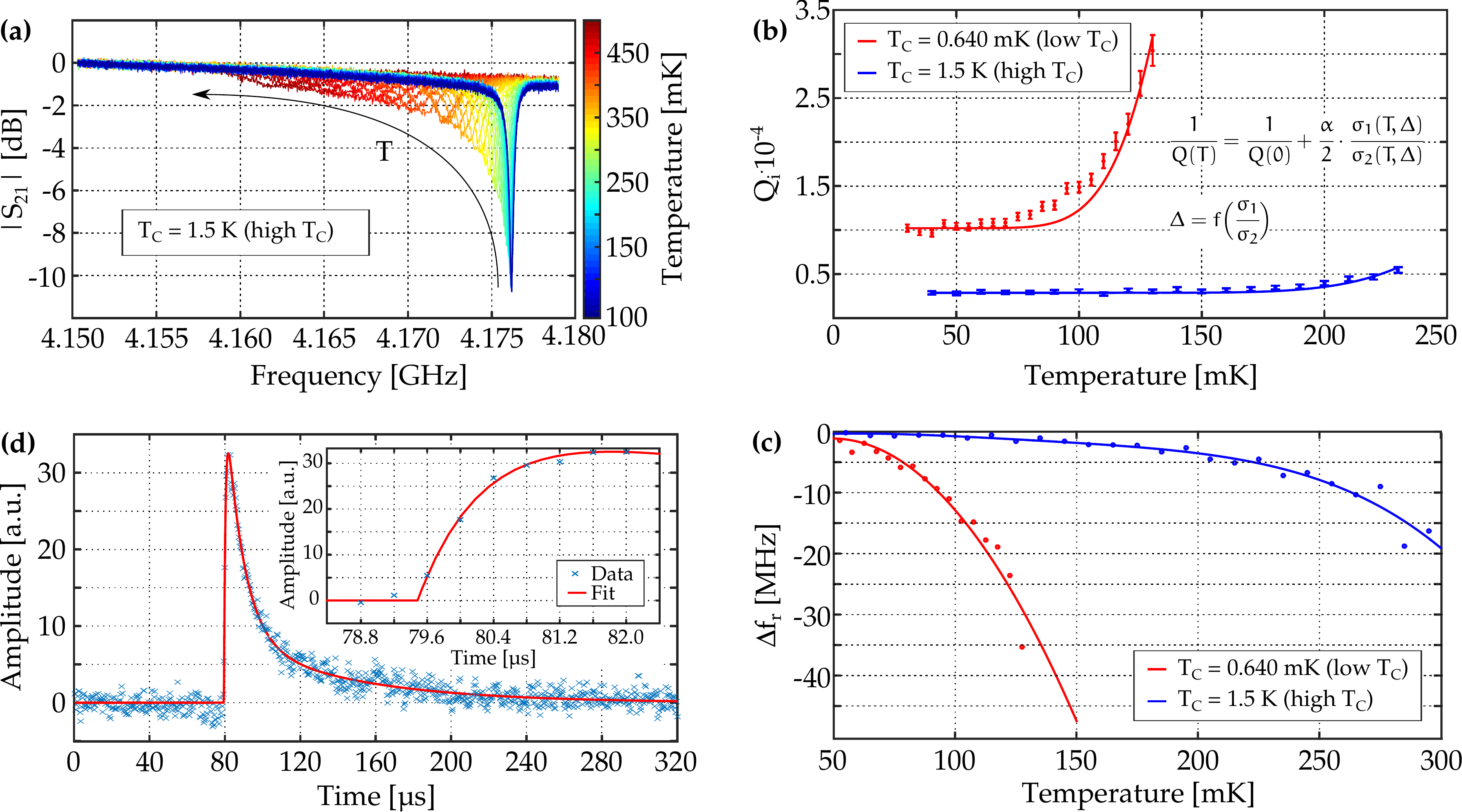}
  \end{center}
  \caption{\label{fig:VsTemp} (a) Resonance characterization as a function of the temperature and extrapolation of the internal quality factors $Q_i$; (b) extrapolation of the energy gap $\Delta$; (c) decrease of resonance frequency $f_r$ with increasing temperature $T$. The variation is steeper with lower critical temperatures; (d) example of observed microresonator response for the produced Ti/TiN multilayers. Data are digitized at 2.5\,MS/s and projected into the frequency and dissipation quadrature.}
\end{figure}

The produced chips were placed in a dilution fridge (Oxford MX 40) and cooled down to temperatures as low as 100~mK. For each resonator, we estimated the critical temperature ($T_c$), the resonance frequency ($f_0$) and the quality factor ($Q_i$) by measuring and fitting the forward transmission $S_{21}$ as a function of the temperature, figure \ref{fig:VsTemp} (a). For each temperature value, we estimated the internal quality factor $Q_i$ and, fitting its behavior versus the temperature, it was possible to estimate the energy gap $\Delta$, according to the Mattis-Bardeen theory~\cite{MattisBardeen}, figure \ref{fig:VsTemp} (b). Characterization results for the first production are shown in table \ref{tab:TiN}. The films show a value of $0.26-0.95$ for the kinetic inductance fraction $\alpha$, which corresponds to an increase by more than one order of magnitude, compared with microresonators built in Aluminum with the same thickness. The \textit{low $T_c$} samples showed a high fraction of kinetic inductance $\alpha$ but a lower quality factor $Q$, while the \textit{high $T_c$} samples showed an energy gap three times lower than the Tantalum one, as needed, but a lower kinetic fraction with respect to the values achieved with Ti/TiN multilayers with the same $T_c$ in a previous production. This is probably due to a not optimized space between the ground planes and the central inductor in the new design geometry. As supposed, the variation of the resonant frequency as a function of the temperature is steeper with lower critical temperatures, figure \ref{fig:VsTemp} (c). These results demonstrate that the proposed technology solutions can satisfy the requirements but, for further developments, optimized designs are needed.

\begin{table}[!t]
{\footnotesize
\begin{center}
\begin{tabular}{ccccccc}
\toprule
Ti    & TiN    & N layers  & $T_c$ & $\Delta$ & $\alpha$&  $Q_i$\\
{[nm]}& {[nm]} &           & [K]   & {[meV]}  &         &      \\
\midrule
10 & 12 & 9  & 1.5 & $0.200\pm 0.004 $ & $0.26\pm 0.01$  & $<10^5$\\   
10 & 7  & 12 & 0.6 & $0.091\pm 0.001 $ & $0.95\pm 0.01 $ & $<10^4$\\
\bottomrule
\end{tabular}
\end{center}}
\caption{\label{tab:TiN} Characterization results from two different families of chips with two different critical temperatures, low $T_c=640$\,mK and high $T_c=1.5$\,K~\cite{FAVERZANI_PHD}.}
\end{table}

Despite that the produced microresonators were not provided with an energy absorber, we irradiated the sensitive part (the inductor) with radioactive sources (\textsuperscript{55}Fe and \textsuperscript{241}Am+Al foil) with the goal to characterize the the detector response time. Measurements showed a typical value of 10\% to 90\% rise time of $\tau_{10-90}=1.6\,\mu$s that corresponds at an exponential rise time lower than $\tau=1\,\mu$s, figure \ref{fig:VsTemp} (d). We obtained similar values for the two produced families. In fact, the rise time it is not correlated with the critical temperature but with the resonator geometry design that, in both cases, was designed in order to have a value around $1\,\mu$s.       

\begin{figure}[!t] 
 \begin{center}
    \includegraphics[clip=true,width=\textwidth]{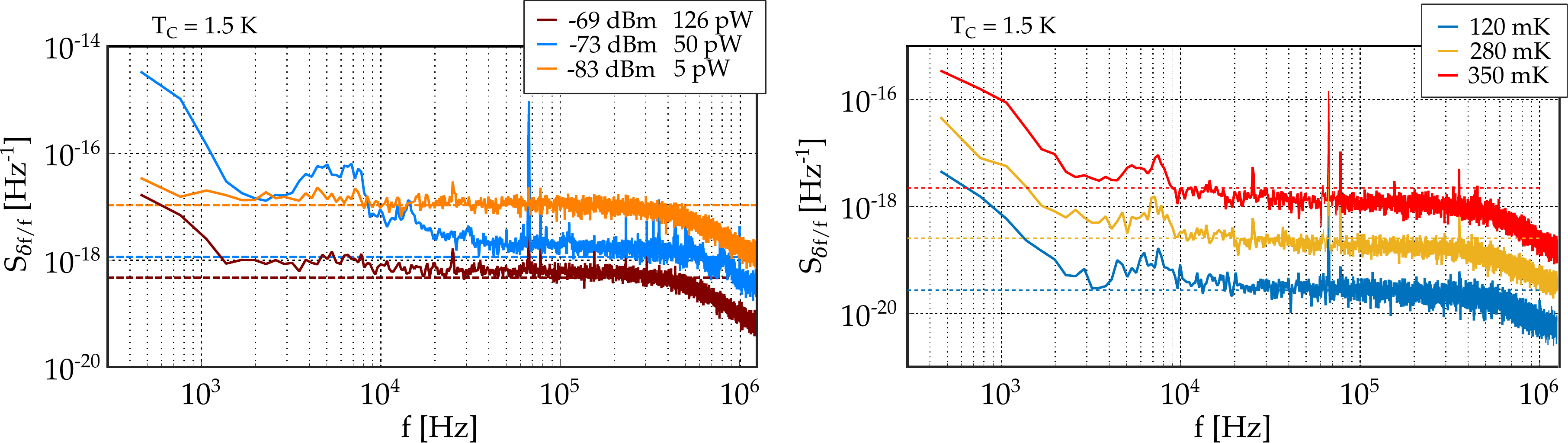}
  \end{center}
  \caption{\label{fig:noise} Noise power spectrum of a Ti/TiN 10/12 chip measured as a function of the driving power (right) and as a function of the temperature (left)~\cite{FAVERZANI_PHD}. The dashed lines are the expected noise level set by the HEMT amplifier. In the spectra the roll-off due to the anti-aliasing filter at 500\,kHz is visible.}
\end{figure}

To evaluate the noise of the produced resonators, we acquired long baselines and performed a rejection trigger to avoid pulse events. The trigger threshold was set just above the noise maximum value. The obtained spectra were compared with the expected noise caused by the HEMT amplifier. In all the cases, the noise resulted to be limited by the HEMT, with an excess noise observed at low frequency. The shape of the low frequency noise does not change when varying the temperature (figure \ref{fig:noise} right) and the driving power (figure \ref{fig:noise} left). For this reason, the excess noise at low frequencies was not identified as the contribution of the Two Level Systems noise (TLSs), but as due to the presence of pulses below the trigger threshold. In fact, given the small thickness of the Ti/TiN film, the majority of the x-rays interact in the substrate, giving rise to a position-dependent response~\cite{Faverzani_LTD15}. For this reason, the pulses response resulted distributed in the whole energy interval ranging between $2\Delta$ up to the source energy. Many pulses can lie in the energy region where they are not identified as pulses because their amplitude is comparable or lower than the intrinsic noise of the amplifier. In this situation, they contribute to the low frequency portion of the noise spectrum. Given the relatively high interaction rate of the X-rays with the substrate, it was not possible to acquire pulse-free baselines long enough to resolve the low frequency region.


\section{Conclusion}
We fabricated and characterized superconducting films in two different families measuring all the relevant parameters. Results prove that the superconductive transition for Ti/TiN multilayers can be tuned in the temperature range from 70\,mK to 4.6\,K by choosing properly the Ti and TiN thicknesses with a good reproducibility and uniformity. The \textit{high} $T_c$ devices satisfied all the design requirements except for a low kinetic inductance fraction $\alpha$, probably due to a not optimized design geometries. The \textit{low} $T_c$ devices presented a steeper variation of the resonant frequency as a function of the temperature, as required for the thermal mode, but lower internal quality factors. At last, the measured rise time was close to the designed value of $1\,\mu$s for both the families. Considering these results, the future plans are the optimization of the microresonator geometry in order to improve the kinetic inductance fraction also in case of high $T_c$ and, since the internal quality factor worsens with lower critical temperature, to find a compromise between the resonant frequency variation and internal quality factor in the case of low $T_c$ device. Finally, the absorber coupled microresonator will be implemented.

\begin{acknowledgements}
This work was supported by the European Research Council (FP7/2007-2013) under Grant Agreement HOLMES no. 340321 and by Fondazione Cariplo through the project \textit{Development of Microresonator Detectors for Neutrino Physics} (grant \textit{International Recruitment Call  2010}, ref. 2010-2351). We also acknowledge the support from INFN through the MARE project. 
\end{acknowledgements}


\begin{thebibliography}{99}

\bibitem{HOLMES}
B. Alpert \textit{et al.}, {\it The European Physical Journal C} \textbf{75}; 112 (2015), 

\bibitem{DeRujula}
A. D. Rujula and M. Lusignoli, {\it Physics Letters B} \textbf{118}, 429–434 (1982); 

\bibitem{PENNING}
S. Eliseev \textit{et al.}, {\it Physical Review Letters} \textbf{115}, 062501 (2015); 

\bibitem{NUCCIOTTI}
A. Nucciotti, {\it The European Physical Journal C} \textbf{74} 3161 (2014); 

\bibitem{NuMECS}
M.P. Croce \textit{et al.}, \textit{arXiv:1510.03874 [physics.ins-det]} (2015);

\bibitem{ECHo}
L. Gastaldo \textit{et al.}, {\it J. Low Temp. Phys.} \textbf{176} 876-884 (2014);

\bibitem{Enss}
A. Fleischmann \textit{et al.}, {\it Topics in Applied Physics} \textbf{99} 151-216 (2005); 

\bibitem{TDM}
C. A. Kilbourne \textit{et al.}, {\it Proc. SPIE} \textbf{7011} 701104–701104–12 (2008);

\bibitem{CDM}
J. Ullom \textit{et al.}, {\it IEEE Trans. Appl. Supercond.} \textbf{13} 643–648 (2003);

\bibitem{FDM}
G. M. Stiehl \textit{et al.}, {\it  Appl. Phys. Lett.} \textbf{100} 072601 (2012);

\bibitem{RFSQUID}
O. Noroozian \textit{et al.}, {\it Appl. Phys. Lett.} \textbf{103} 202602 (2013);

\bibitem{ARCONS}
B. A. Mazin, \textit{et al.}, PASP \textbf{125}, 1348-1361 (2013); 

\bibitem{GIGAZ}
D. W. Marsden \textit{et al.}, ApJS, \textbf{208}, 208, (2013); 

\bibitem{KRAKENS}
B. A. Mazin, \textit{et al.}, Keck SSC Whitepaper (2015);

\bibitem{Day}
P. K. Day \textit{et al.}, {\it Nature} \textbf{425}, 817-821 (2003); 

\bibitem{Tantalum}
B. A. Mazin \textit{et al.}, {\it Appl. Phys. Lett.} \textbf{89}, 222507; 

\bibitem{ThermalEqui}
J. Gao \textit{et al.}, {\it J. Low Temp. Phys.}, \textbf{151}, 557–563 (2008); 

\bibitem{TKID}
G. Ulbricht \textit{et al.}, {\it Appl. Phys. Lett.} \textbf{106}, 251103 (2015); 

\bibitem{Faverzani_LTD15}
M. Faverzani \textit{et al.}, {\it J. Low Temp. Phys.} \textbf{176}, 530-537, (2014), 

\bibitem{Multilayer}
A. Giachero \textit{et al.}, {\it J. Low Temp. Phys.}, \textbf{175}, 155-160 (2014); 

\bibitem{PROXIMITY}
W. Silvert, {\it J. Low Temp. Phys.}, \textbf{20}, 439–477 (1975); 

\bibitem{MattisBardeen}
D. C. Mattis and J. Bardeen, {\it Physical Review}, \textbf{111}, 412–417 (1958); 

\bibitem{FAVERZANI_PHD}
M. Faverzani, Ph.D. Thesis (2015), University of Milano-Bicocca;







\end{thebibliography}
\end{document}